\documentclass[runningheads]{llncs}
\usepackage[colorlinks=true]{hyperref}
\usepackage{graphicx}
\usepackage{multirow}
\usepackage{tikz}
\usetikzlibrary{arrows,automata,shapes,decorations,patterns,positioning,graphs} %,graphdrawing}
\usepackage{forest}
\usepackage{amsmath}
\usepackage{amsfonts}
\usepackage{amssymb}
\usepackage{relsize}
\usepackage{mathtools}
\usepackage{algorithm}
\usepackage{algcompatible}
\usepackage{soul}
\usepackage{listings}

\lstdefinelanguage{Prism} {
    keywords=[0]{const,global,bool,int,double,dtmc,ctmc,mdp,csg,rewards,endrewards,formula,module,endmodule,player,endplayer,init,true,false},
    basicstyle=\linespread{1.0}\scriptsize\ttfamily\color{black},
    keywordstyle=[0]\ttfamily\color{black},
    comment=[l][\color{gray}]{//},
    sensitive=true,
    frame=single,	
    numbers=left,
    numberstyle=\tiny\color{gray},
    morestring=[b]"
}

\algnewcommand{\PROCEDURE}[1]{\STATE \textbf{procedure}\space #1}
\algnewcommand{\ENDPROCEDURE}{\STATE \textbf{end procedure}}
\algnewcommand\RETURN{\STATE \textbf{return}\space}
\algnewcommand\TRUE{\textbf{true}}
\algnewcommand\FALSE{\textbf{false}}
\algnewcommand\OR{\textbf{or}\space}
\algnewcommand\AND{\textbf{and}\space}
\algnewcommand{\IFTHEN}[2]{\STATE \algorithmicif\ #1\ \algorithmicthen\ #2\ \algorithmicend\ \algorithmicif}
\algnewcommand{\IFTHENELSE}[3]{\STATE \algorithmicif\ #1\ \algorithmicthen\ #2\ \algorithmicelse\ #3 \algorithmicend\ \algorithmicif}

\newcommand{\uu}{\underline{\hspace{0.4em}}}
\newcommand{\best}[1]{{\bf{#1}}}

\colorlet{grayhl}{gray!20}
\sethlcolor{grayhl}

\tikzset{
  0 my edge/.style={densely dashed, my edge},
  my edge/.style={},
}
\forestset{
  BDT/.style={
    for tree={
      if n children=0{}{circle},% use a circle unless there are 0 children
      draw,% draw every node
      edge={
        my edge,% use the my edge style for edges (with the arrow)
      },
      if n=1{
        edge+={0 my edge},% if the child is the first one, add the 0 my edge style (dashed)
      }{},
      font=\sffamily,% use sans serif for node text
    }
  },
}

\newcommand\encfunc[1]{{\sf enc}_{#1}}
\newcommand\enc[2]{{\encfunc{#1}({#2})}}
\newcommand\dd[1]{{{\sf #1}}}
\newcommand\ddvars[1]{{\underline{\sf #1}}}
\newcommand\ddfunc[1]{{f_{\sf #1}}}
\newcommand\ddfuncs[2]{{f_{{\sf #1}_{#2}}}}
\newcommand\ddop[1]{{\mbox{\sc #1}}}
\newcommand\alg[1]{{\mbox{\sc #1}}}
\newcommand\valn[1]{{\underline{#1}}}

\renewcommand{\emptyset}{\varnothing}

\DeclareMathOperator{\opt}{opt}

\usepackage{scalerel}

\newcommand{\sectref}[1]{Section~\ref{#1}}

\newcommand{\figref}[1]{Figure~\ref{#1}}
\newcommand{\tabref}[1]{Table~\ref{#1}}

\newcommand{\algoref}[1]{Algorithm~\ref{#1}}

\spnewtheorem{assumption}{Assumption}{\bfseries}{\itshape}
\newcounter{exampcount}
\newenvironment{examp}
{\refstepcounter{exampcount}
\vskip6pt\noindent
{\bf Example \arabic{exampcount}.}}
{\hfill$\blacksquare$\vskip6pt}

\newcommand{\startpara}[1]{{%
\vskip6pt\noindent
{\bf #1.}}}

\def\matr#1{{\mathbf{#1}}}
\def\vect#1{{\underline{#1}}}

\def\dd#1{{\mathsf{#1}}}

\def\ra{{\rightarrow}}

\def\cC{{\mathcal{C}}}

\def\Bset{\mathbb{B}}
\def\Nset{\mathbb{N}}

\def\Rset{\mathbb{R}}

\def\Eset{\mathbb{E}}
\def\Rsetgeq{\mathbb{R}_{\geq 0}}

\def\ra{\rightarrow} % right arrow
\def\rmdef{\,\stackrel{\mbox{\rm {\tiny def}}}{=}}
\newcommand{\true}{\mathtt{true}} % logical true
\newcommand{\false}{\mathtt{false}} % logical false

\renewcommand{\leq}{\leqslant}
\renewcommand{\geq}{\geqslant}

\newcommand\game{{\mathcal G}}

\newcommand\sinit{{\bar{s}}}

\newcommand\Act{\act}

\newcommand\dist{{\mathit{Dist}}}
\newcommand\Dist{{\dist}}

\newcommand\Prob{{\mathit{Prob}}}

\newcommand\val{{\mathit{val}}}

\newcommand{\rewfunc}{\mathit{rew}} %rewards function
\newcommand{\last}{\mathit{last}}
\newcommand{\ipaths}{\mathit{IPaths}}
\newcommand{\fpaths}{\mathit{FPaths}}

\def\sateps{{\,\models\,}}
\def\notsateps{{\,\not\models\,}}

\def\Sat{{\mathit{Sat}}}
\newcommand{\coalition}[1]{\langle \! \langle {#1} \rangle \! \rangle}

\def\next{{X\,}}
\def\until{{\ {\cal U}\ }}
\def\buntil{{\ {\cal U}^{\leq k}\ }}

\def\Act{{\mathit{Act}}}

\def\next{{\mathtt X\,}}
\def\until{{\ \mathtt{U}\ }}

\def\buntil{{\ \mathtt{U}^{\leq k}\ }}

\def\future{{\mathtt{F}\ }}
\def\futureop{{\mathtt{F}}}

\newcommand\bfuturep[1]{{\mathtt{F}^{\leq #1}\ }}
\def\globally{{\mathtt{G}\ }}

\newcommand{\scumul}[1]{\mathtt{C}^{#1}} %{\leq bound}
\newcommand{\ap}{\mathsf{a}}
\newcommand{\sinstant}[1]{\mathtt{I}^{#1}} %{= bound}
\newcommand\probopP{{\mathtt P}}
\newcommand\nashop[3]{\coalition{#1}_{#2}(#3)}

\newcommand\probop[2]{\probopP_{#1}[\,{#2}\,]}

\newcommand\rewopR{{\mathtt R}}
\newcommand\rewop[3]{\rewopR^{#1}_{#2}[\,{#3}\,]}

\newcommand{\lab}{{\mathit{L}}}

\newcommand{\rew}{\mathit{r}}

\begin{document}
\title{Symbolic Verification and Strategy Synthesis \\ for Turn-based Stochastic Games}
\titlerunning{Symbolic Verification for Turn-based Stochastic Games}
\author{Marta Kwiatkowska\inst{1} \and Gethin Norman\inst{1,2} \and David~Parker\inst{1,3} \and Gabriel Santos\inst{1}}
\institute{Department of Computing Science, University of Oxford, UK
\and
School of Computing Science, University of Glasgow, UK
\and School of Computer Science, University of Birmingham, UK}
\maketitle 
\begin{abstract}
Stochastic games are a convenient formalism for modelling systems that comprise
rational agents competing or collaborating within uncertain environments.
Probabilistic model checking techniques for this class of models
allow us to formally specify quantitative specifications of either collective or individual
behaviour and then automatically synthesise strategies for the agents
under which these specifications are guaranteed to be satisfied.
Although good progress has been made on algorithms and tool support,
efficiency and scalability remain a challenge.
In this paper, we investigate a symbolic implementation
based on multi-terminal binary decision diagrams.
We describe how to build and verify turn-based stochastic games
against either zero-sum or Nash equilibrium based temporal logic specifications.
We collate a set of benchmarks for this class of games,
and evaluate the performance of our approach, showing that it
is superior in a number of cases and that strategies synthesised
in a symbolic fashion can be considerably more compact.
%\keywords{First keyword  \and Second keyword \and Another keyword.}
\end{abstract}

\section{Introduction}\label{sec:intro}

Games have long been used as an underlying modelling formalism
for the design and verification of computerised systems.
For example, they naturally model the interaction between a system,
whose behaviour can be controlled, and its environment whose actions cannot.
Another example is the interplay between the defender and attacker in a computer security scenario.

In the context of \emph{model checking}, where the required behaviour of a system is
specified using temporal logic, we can use, for example,
alternating-time temporal logic (ATL)~\cite{AHK02}
to formalise the capabilities of a player (or a coalition of players)
acting in the context of another, adversarial player (or coalition) in a game model.
Yet further expressive logics such as strategy logic~\cite{CHP10}
can also reason about the existence of, for instance, Nash equilibria.

Another important tool for modelling and verification is \emph{stochasticity}.
Probability is often essential to effectively quantify uncertain aspects of systems,
from the presence of hardware failures to the unreliability of physical sensors.
\emph{Stochastic games}~\cite{Sha53,Con92,FV97} are a well studied model
for the dynamic execution of multiple players in a probabilistic setting.
Results and algorithms for many verification problems on such
models have also been presented, e.g.,~\cite{CJH04,KH12}.

Building on these foundations, progress has since been made on the
practical applicability of probabilistic model checking using stochastic games.
This includes logics and algorithms for both \emph{turn-based} stochastic games (TSGs)~\cite{CFK+13b}
and \emph{concurrent} stochastic games (CSGs)~\cite{KNPS21}.
The logic rPATL, a quantitative extension of ATL,
allows specification of zero-sum properties for stochastic games,
and extensions~\cite{KNPS21} also permit reasoning about the existence of Nash equilibria.
A modelling formalism and tool support for TSGs and CSGs
have been developed, in the form of PRISM-games~\cite{KNPS20},
and this framework has been successfully applied to the analysis of, for example,
human-robot collaborations~\cite{FWHT16,JJK+18},
self-adaptive software systems~\cite{CGSP15}
and computer security~\cite{ANP16}.

However, as usual for model checking approaches,
efficiency and scalability are key challenges.
So, in this paper, we consider \emph{symbolic} implementations,
in particular using binary decision diagrams (BDDs)
and multi-terminal BDDs (MTBDDs), previously deployed
for the compact representation and efficient manipulation of various models.
Well known tools for verifying multi-agent systems
such as MOCHA~\cite{AHM+98} and MCMAS~\cite{LQR09}
incorporate symbolic implementations of model checking,
and probabilistic model checkers such as PRISM~\cite{KNP11} and STORM~\cite{DJKV17}
support symbolic techniques for simpler classes of stochastic models,
such as Markov chains and Markov decision processes (MDPs).

As a first step in this direction,
we consider a symbolic implementation of model checking and strategy synthesis
for \emph{turn-based} stochastic games.
This also provides symbolic verification of (turn-based) probabilistic timed games,
via the digital clocks translation~\cite{KNP19}.
We describe how to encode TSGs as MTBDDs
and how to perform verification symbolically,
in particular using value iteration.
We also describe how to perform strategy synthesis,
and how to extend this approach to compute Nash equilibria for TSGs.

In order to evaluate this,
we collate a set of TSG model checking benchmarks of varying sizes,
and add them to the PRISM benchmark suite~\cite{KNP12b}.
We show that the symbolic approach offers significant gains in terms of
the time required for model construction, for qualitative (graph-based) verification
and, in some cases, for numerical solution of TSGs.
We also show that optimal strategies can be represented more compactly
symbolically rather than explicitly.

\startpara{Related work}
Various methods have been proposed for solving stochastic games~\cite{Sha53,Con92,FV97}
and for verifying them against logical specifications, e.g.,~\cite{CJH04,KH12}.
GIST~\cite{CHJR10} implements qualitative verification against $\omega$-regular specifications
and PRISM-games~\cite{KNPS20} supports various quantitative properties.
In \cite{KRSW20}, a wider range of methods for solving TSGs are implemented and explored,
offering significant speed-ups. However, none of these provide symbolic implementations.

Multiple MTBDD-based implementations of probabilistic model checking
for simpler (non-game) stochastic models have been developed.
Originally, this focused on PCTL model checking for Markov chains or MDPs~\cite{BCHG+97,Bai98,Par02}.
A so-called \emph{hybrid} approach~\cite{Par02} improves performance
through a combination of symbolic model storage and explicit-state algorithms,
and is the default model checking engine in PRISM.
Enhancements by others include automatic variable reordering
and model checking of quantile-based properties~\cite{KBC+16}.
MTBDDs and BDDs have also been applied to the solution of
energy games~\cite{AMPR21}, which are \emph{non-probabilistic} games with integer weights.
Extensions of MTBDDs (XADDs) have been used for symbolic analysis of
continuous-state MDPs~\cite{ZSC12}.

Interpreting ``symbolic'' verification more widely, i.e., beyond BDD-based approaches,
\cite{MMKS21} considers symbolic methods for stochastic parity games,
and \cite{BJKK+20} presents a probabilistic variant of the well known IC3 approach to model checking.
Also relevant are methods to use, and learn, decision trees to succinctly represent strategies for probabilistic models~\cite{BCC+15}.

\section{Preliminaries}\label{sec:prelim}

We begin with some background material,
first on model checking of turn-based stochastic games,
and then on (multi-terminal) binary decision diagrams.

\startpara{Notation}
We use $\dist(X)$ to denote the set of probability distributions over a set $X$,
and we use $\Bset=\{0,1\}$ for the set of Boolean values,
with 0 denoting $\false$ and 1 denoting $\true$.

\subsection{Model Checking for Stochastic Games}\label{sec:bg_mc}

Several variants of stochastic games exist.
In this paper, we focus on (finite, multi-player) turn-based stochastic games.

\begin{definition}[Turn-based stochastic game] \label{def:tsg}
A \emph{turn-based stochastic game} (TSG) is a tuple $\game = (N, S, (S_i)_{i \in N}, \sinit, A, \delta, \lab)$ where:

\begin{itemize}
\item $N$ is a finite set of players;
\item $S$ is a finite set of states;
\item $(S_i)_{i \in N}$ is a partition of $S$;
\item $\sinit \in S$ is an initial state;
\item $A$ is a finite set of actions;
\item $\delta : S \times A \rightarrow Dist(S)$ is a (partial) transition probability function;
\item $\lab : S \rightarrow 2^{AP}$ is a labelling function.
\end{itemize}
\end{definition}
We now fix an $n$-player TSG $\game$ for the remainder of the section.
The TSG $\game$ starts in its initial state $\sinit \in S$. In each state $s$, a player $i \in N$ selects an action from the set of \emph{available} actions, which is denoted by $A(s) = \{a \in A\ |\ \delta(s,a) \; \mbox{is defined} \}$. We assume that $A(s) \neq \varnothing$ for all states $s$. The choice of action to take in each state $s$ is under the control of exactly one player, namely the player $i \in N$ for which $s \in S_i$. Once action $a \in A(s)$ is selected, the successor state is chosen according to the probability distribution $\delta(s, a)$, i.e., the game moves to state $s'$ with probability $\delta(s,a)(s'$). We augment $\game$ with \emph{reward structures}, which are tuples of the form $r=(r_A,r_S)$ where $r_A : S {\times} A \ra \Rset$ and $r_S : S \ra \Rset$ are action and state reward functions, respectively.

A \emph{path} through $\game$ is a sequence $\pi = s_0 \xrightarrow{a_0} s_1 \xrightarrow{a_1} \cdots$ such that $s_i \in S$, $a_i \in A(s_i)$ and $\delta(s_i,a_i)(s_{i+1}) > 0$ for all $i \geq 0$. %Given a path $\pi$, we denote by $\pi(i)$ the $(i{+}1)$th state, $\pi[i]$ the $(i{+}1)$th action, and if $\pi$ is finite, $\last(\pi)$ the final state. %
The sets of finite and infinite paths (starting in state $s$) of $\game$ are given by $\fpaths_\game$ and $\ipaths_\game$ ($\fpaths_{\game,s}$ and $\ipaths_{\game,s}$).

\emph{Strategies} of $\game$ are used to resolve the choices of the players. Formally, a strategy for player $i$ is a function $\sigma_i \colon \fpaths_{\game} \ra \dist(A)$ such that, if $\sigma_i(\pi)(a_i){>}0$, then $a_i \in A(\last(\pi))$ where $\last(\pi)$ is the final state of path $\pi$. A \emph{strategy profile} is a tuple $\sigma = (\sigma_1,\dots,\sigma_{n})$ of strategies for all players. The set of strategies for player $i$ and set of profiles are denoted $\Sigma^i_\game$ and $\Sigma_\game$. Given a profile $\sigma$ and state $s$, let $\ipaths^\sigma_{\game,s}$ denote the infinite paths with initial state $s$ corresponding to $\sigma$.
We can then define, using standard techniques~\cite{KSK76}, a probability measure $\Prob^{\sigma}_{\game,s}$ over $\ipaths^{\sigma}_{\game,s}$ and, for a random variable $X \colon \ipaths_{\game} \rightarrow \Rset$, the expected value $\Eset^{\sigma}_{\game,s}(X)$ of $X$ from $s$ under $\sigma$.

In $\game$, the utility  or \emph{objective} of player $i$ is represented by a random variable $X_i \colon \ipaths_{\game} \rightarrow \Rset$. Such variables can encode, for example, the probability of reaching a target or the expected cumulative reward before reaching a target.

We  now introduce the notion of Nash equilibrium (NE)~\cite{NMK+44} for $\game$ given objectives $( X_i )_{i=1}^n$ for the players. We restrict our attention to subgame-perfect NE~\cite{OR04}, which are NE in every state of $\game$. For profile $\sigma=(\sigma_1,\dots,\sigma_n)$ and player $i$ strategy $\sigma_i'$, we define the sequence $\sigma_{-i} \rmdef (\sigma_1,\dots,\sigma_{i-1},\sigma_{i+1},\dots,\sigma_n)$ and profile $\sigma_{-i}[\sigma_i'] \rmdef (\sigma_1,\dots,\sigma_{i-1},\sigma_i',\sigma_{i+1},\dots,\sigma_n)$.
\begin{definition}[Best response] 
For objectives $( X_i )_{i=1}^n$, player $i$, strategy sequence $\sigma_{-i}$ and state $s$, 
a \emph{best response} for player $i$ to $\sigma_{-i}$ in state $s$ is a strategy $\sigma^\star_i$ for player $i$ such that $\Eset^{\sigma_{-i}[\sigma^\star_i]}_{\game,s}(X_i) \geq \Eset^{\sigma_{-i}[\sigma_i]}_{\game,s}(X_i)$ for all $\sigma_i \in \Sigma^i_\game$.
\end{definition}
\begin{definition}[Nash equilibrium]\label{nash-def}
For objectives $( X_i )_{i=1}^n$, a strategy profile $\sigma^\star=(\sigma^\star_1,\dots,\sigma^\star_n)$ of $\game$ is a subgame-perfect \emph{Nash equilibrium} (NE) if $\sigma_i^\star$ is a best response to $\sigma_{-i}^\star$ for all $i \in N$ and $s \in S$.
Furthermore, a NE $\sigma^\star$ of $\game$ is a \emph{social welfare optimal NE} (SWNE) for objectives $X_1,\dots,X_n$  if  $\Eset^{\sigma^\star}_{\game,s}(X_1){+}\cdots$ ${+} \Eset^{\sigma^\star}_{\game,s}(X_n) \geq \Eset^{\sigma}_{\game,s}(X_1) {+} \cdots+ \Eset^{\sigma}_{\game,s}(X_n)$ for all NE $\sigma$ of $\game$.
\end{definition}
We can also define the dual concept of \emph{social cost optimal NE} (SCNE)~\cite{KNPS21}, for which the players of $\game$ try to minimise, rather than maximise, their expected utilities by considering equilibria for the objectives $-X_1,\dots,-X_n$.

To formally specify properties of TSGs, we use the PRISM-games logic presented in~\cite{KNPS21}, which extends the logic rPATL previously defined for zero-sum properties of TSGs~\cite{CFK+13b}.
The logic uses the \emph{coalition} operator $\coalition{C}$ from alternating temporal logic (ATL)~\cite{AHK02} to define \emph{zero-sum} formulae and allows \emph{nonzero-sum} properties, using (social welfare or social cost) NE.

\begin{definition}[PRISM-games logic syntax]\label{def:rpatlsyntax}
The syntax of the PRISM-games logic is given by the grammar:
\begin{eqnarray*}
\phi & \; \coloneqq \; & \mathtt{true} \mid \ap \mid \neg \phi \mid \phi \wedge \phi \mid \coalition{C}\probop{\bowtie p}{\psi} \mid \coalition{C}\rewop{r}{\bowtie q}{\rho} \mid \nashop{C_{1}{:}\cdots{:}C_m}{\opt \bowtie q}{\theta} \\
\psi & \; \coloneqq \; & \next \phi \ \mid \ \phi \buntil \phi \ \mid \ \phi \until \phi \\
\rho & \; \coloneqq \; & \sinstant{=k} \ \mid \ \scumul{\leq k} 
\ \mid \  \future \phi \\
\theta & \; \coloneqq \; &\probop{}{\psi}{+}{\cdots}{+}\probop{}{\psi} \ \mid \  \rewop{r}{}{\rho}{+}{\cdots}{+}\rewop{r}{}{\rho} 
\end{eqnarray*}
where $C$ and $C_1,,\dots,C_m$ are coalitions of players such that $C_i \cap C_j = \emptyset$ for all $1\leq i \neq j \leq m$ and $\cup_{i=1}^m C_i = N$, $\bowtie\,\in\!\{<,\leq,\geq,>\}$, $p \in [0,1] \cap \Rset$, $\rew$ is a reward structure, $q \in \Rsetgeq$, $\opt \in \{ \min,\max\}$, $\ap$ is an atomic proposition and $k \in \Nset$. %
\end{definition}
The syntax of the PRISM-games logic distinguishes between state ($\phi$), path ($\psi$), reward ($\rho$) and nonzero-sum ($\theta$) formulae. State formulae are evaluated over states of a TSG, while path, reward and nonzero-sum formulae are evaluated over paths.
\emph{Zero-sum} state formula have the following meaning:
\begin{itemize}
\item
$\coalition{C} \probop{\bowtie q}{\psi}$ is satisfied in a state if the coalition of players $C$ can ensure that the probability of the path formula $\psi$ being satisfied is ${\bowtie}\,q$, regardless of the actions of the other players;
\item 
$\coalition{C} \rewop{r}{\bowtie x}{\rho}$ is satisfied in a state if the players in $C$ can ensure that the expected value of the reward formula $\rho$ for reward structure $r$ is ${\bowtie}\,x$, regardless of the actions of the other players.
\end{itemize}
On the other hand, for a \emph{nonzero-sum} state formula:
\begin{itemize}
\item
$\nashop{C_1{:}{\cdots}{:}C_m}{\max\bowtie x}{\theta}$ is satisfied
if there exists a subgame-perfect SWNE profile between coalitions $C_1,\dots,C_m$ under which the \emph{sum} of the objectives of $C_1,\dots,C_m$ in $\theta$ is ${\bowtie}\,x$;
\item
$\nashop{C_1{:}{\cdots}{:}C_m}{\min\bowtie x}{\theta}$ is satisfied
if there exists a subgame-perfect SCNE profile between coalitions $C_1,\dots,C_m$ under which the \emph{sum} of the objectives of $C_1,\dots,C_m$ in $\theta$ is ${\bowtie}\,x$.
\end{itemize}
For all of the above formulae, we also allow \emph{numerical} variants,
which directly yield an optimal value, rather than checking whether a threshold can be met.
For example, $\coalition{C} \probop{\max=?}{\psi}$ gives the maximum probability
with which the players in $C$ can guarantee that that $\psi$ is satisfied.

Both zero-sum and nonzero-sum formulae
are composed of \emph{path} ($\psi$) and \emph{reward} ($\rho$) formulae,
used in the probabilistic and reward objectives
included within $\probopP$ and $\rewopR$ operators, respectively.
The path formulae include: \emph{next} ($\next \phi$), \emph{bounded until} ($\phi \buntil \phi$) and \emph{unbounded until} ($\phi \until \phi$).
There are also the standard equivalences including: \emph{probabilistic reachability} ($\future\phi \equiv \true\until\phi$) and \emph{bounded probabilistic reachability} ($\bfuturep{k}\phi \equiv \true\buntil\phi$). 

The reward formulae include: \emph{instantaneous (state) reward} at the $k$th step ($\sinstant{=k}$),
\emph{bounded cumulative reward} over $k$ steps ($\scumul{\leq k}$),
and \emph{reachability reward} until a formula $\phi$ is satisfied ($\future \phi$).
In the case of reachability reward formulae, several variants have previously been introduced~\cite{CFK+13b},
differing in how they treat paths that do not reach a state satisfying $\phi$.
%For simplicity,
We restrict our attention to the most common one,
the default in PRISM, which assigns the reward value infinity to paths that never reach a state satisfying $\phi$. 

We next define the semantics of the PRISM-games logic for TSGs. However, we first need to define the concept of a coalition game.
\begin{definition}[Coalition game] For TSG $\game$ %=(N, S, (S_i)_{i \in N}, \sinit, A, \delta, \lab)$
and a partition of its players into $m$ coalitions $\cC=\{C_1, \dots, C_{m}\}$,
we define the \emph{coalition game} $\game^\cC = ( \{1,\dots,m\}, S, (S^\cC_i)_{i \in M}, \sinit, A, \delta, \lab)$
as an $m$-player TSG where $S^\cC_i = \cup_{j \in C_i} S_j$.
\end{definition}

To simplify notation, for any coalition $C$ of $\game$, we use the notation $\game^C$ to represent the 2-player coalition game $\game^\cC$ where $\cC = \{C,N{\setminus}C\}$.
\begin{definition}[PRISM-games logic semantics]\label{def:rpatlsemantics}
For a TSG $\game$ and formula $\phi$, we define the satisfaction relation $\sateps$ inductively over the structure of $\phi$.
The propositional logic fragment $(\mathtt{true}$, $\ap$, $\neg$, $\wedge)$ is defined in the usual way.
For a PRISM-games logic formula and state $s \in S$ of TSG $\game$, we have:
\begin{eqnarray*}
s \sateps \coalition{C} \probop{\bowtie q}{\psi}  
& \ \Leftrightarrow \ &
\exists \sigma_1 \in \Sigma^1 . \, \forall \sigma_2 \in \Sigma^2 . \,
\Eset^{\sigma_1,\sigma_2}_{\game^{C},s}(X^\psi) \bowtie q \\
s \sateps \coalition{C} \rewop{r}{\bowtie x}{\rho}
& \ \Leftrightarrow \ &
\exists \sigma_1 \in \Sigma^1 . \, \forall \sigma_2 \in \Sigma^2  . \, \Eset^{\sigma_1,\sigma_2}_{\game^{C},s}(X^{r,\rho}) \bowtie x \\
s \sateps \nashop{C_{1}{:}\cdots{:}C_m}{\opt \bowtie q}{\theta} & \;\; \Leftrightarrow & \;\;
\exists \sigma^\star \in \Sigma_{\game^{\cC}} . \, \big( \, \mbox{$\sum_{i=1}^m$} \Eset^{\sigma^\star}_{\game^{\cC}}(X^\theta_i) \, \big) \bowtie q
\end{eqnarray*}
and $\sigma^\star$
is an SWNE if $\opt = \max$, and an SCNE$\,$ if $\opt = \min$, for the objectives $( X^\theta_i )_{i=1}^m$ in the coalition game $\game^{\cC}$.

For an objective $X^{\psi}$, $X^{r,\rho}$ and path $\pi \in \ipaths_{\game^C,s}:$
\begin{eqnarray*}
X^{\psi}(\pi) & \ = \ & 1 \;\mbox{if $\pi \sateps \psi$ and 0 otherwise} \\
X^{r,\rho}(\pi) & \ = \ & \rewfunc(r,\rho)(\pi) \, .
\end{eqnarray*}
The semantics for satisfaction of path formulae ($\pi \sateps \psi$)
and the random variable $\rewfunc(r,\rho)(\pi)$ for a reward formula
can be found in, e.g.,~\cite{KNPS21}.

\iffalse
For a temporal formula and path $\pi \in \ipaths_{\game^C,s}$:
\begin{eqnarray*}
\pi \sateps \next \phi 
& \ \Leftrightarrow \ & 
\pi(1) \sateps \phi \\
\pi \sateps \phi_1 \buntil \phi_2 
& \ \Leftrightarrow \ &
\exists i \leq k . \, (\pi(i) \sateps \phi_2 \wedge \forall j < i . \, \pi(j) \sateps \phi_1 )
\\
\pi \sateps \phi_1 \until \phi_2 
& \ \Leftrightarrow \ & 
\exists i \in \Nset . \, ( \pi(i) \sateps \phi_2 \wedge \forall j < i  . \, \pi(j) \sateps \phi_1 )
\end{eqnarray*}
For a reward structure $r$, reward formula and path $\pi \in \ipaths_{\game^C,s}$:
\begin{eqnarray*}
\rew(r,\sinstant{=k})(\pi) & \ = \ & r_S(\pi(k)) \\
\rew(r,\scumul{\leq k})(\pi) & \ = \ & \mbox{$\sum_{i=0}^{k-1}$} \big( r_A(\pi(i),\pi[i])+r_S(\pi(i)) \big) \\
\rew(r,\future  \phi)(\pi) & \ = \ & \begin{cases}
\infty
& \mbox{if} \; \forall j \in \Nset . \, \pi(j) \notsateps \phi \\
\mbox{$\sum_{i=0}^{k_\phi-1}$} \big( r_A(\pi(i),\pi[i])+r_S(\pi(i)) \big) & \mbox{otherwise}
\end{cases}
\end{eqnarray*}
where $k_\phi = \min \{ k \mid \pi(k) \sateps \phi \}$.
\fi
\end{definition}
As the zero-sum objectives appearing in the logic are either finite-horizon or infinite-horizon and correspond to either probabilistic until or expected reachability formulae, we have that TSGs are \emph{determined} with respect to these objectives~\cite{Mar98}, which yields the following equivalences:
\[ \begin{array}{rcl}
\coalition{C} \probop{\max=?}{\psi} & \equiv & \coalition{N {\setminus} C} \probop{\min=?}{\psi} \\
\coalition{C} \rewop{r}{\max=?}{\rho} & \equiv & \coalition{N {\setminus} C} \rewop{r}{\min=?}{\rho} \, .
\end{array}
\]
Also, as for other probabilistic temporal logics, we can represent negated path formulae by inverting the probability threshold, e.g.:
\[ \begin{array}{rcl}
\coalition{C} \probop{\geq q}{\neg\psi} &\equiv & \coalition{C} \probop{\leq 1-q}{\psi} \\
% \nashop{C_1{:}{\cdots}{:}C_m}{\max\geq q }{\probop{}{\psi_1}{+}{\cdots}{+}\probop{}{\psi_m}}& \equiv& \nashop{C_1{:}{\cdots}{:}C_m}{\min\leq m-q }{\probop{}{\neg\psi_1}{+}{\cdots}{+}\probop{}{\neg\psi_m}}
\end{array}
\]
notably allowing the `globally' operator $\globally \phi \equiv \neg (\future \neg \phi)$ to be defined.

Since the logic is branching-time, the model checking algorithm for the logic works by recursively computing the set $\Sat(\phi)$ of states satisfying formula $\phi$
over the structure of $\phi$. The main step in the algorithm requires computation of values for zero-sum and nonzero-sum formulae. The standard approach is to use \emph{value iteration}~\cite{CH08}, which we discuss below.

Note that the PRISM logic used here, which includes non-zero sum formulae,
was first considered for CSGs~\cite{KNPS21,KNPS20b}.
Our focus here is on TSGs, where the computation of values is simpler:
because only one coalition has a choice in each state,
value iteration need only take the minimum or maximum over actions,
whereas for CSGs matrix games need to be solved in each state.

For the case of zero-sum formulae, efficiency and accuracy can be improved through the use of graph-based \emph{precomputation algorithms}~\cite{AHK02}, which identify the states that have values 0 and 1 in the case of probabilistic properties and value $\infty$ in the case of expected reward properties.

%\startpara{Precomputation} ...

\startpara{Value iteration} Below, we illustrate value iteration for the zero-sum formula $\phi = \coalition{C} \probop{\max=?}{\future \phi'}$; the remaining cases have a similar structure. The value of $\phi$ in state $s$ is given by the limit $\val(s,\phi) = \lim_{k \ra \infty} x^k_s$, where for any $k \in \Nset$:
\[
x_s^k \ = \  
\begin{cases}
1 & \mbox{if $s \in \Sat(\phi')$} \\ 
0 & \mbox{else if $k=0$} \\ 
\max\limits_{a \in A(s)}\sum_{s'\in S} \delta(s,a)(s') \cdot x_{s'}^{k-1} & \mbox{else if $s \in \cup_{i \in C} S_{i}$} \\ % \in S^{?} \mbox{ and } s \in S_{\Diamond} \\
\min\limits_{a \in A(s)}\sum_{s'\in S} \delta(s,a)(s') \cdot x_{s'}^{k-1} & \mbox{otherwise} % s \in S_{N \setminus C} %\in S^{?} \mbox{ and } s \in S_{\Box}
\end{cases}
\]
In practice, a suitable convergence criterion needs to be chosen to terminate the computation.
Here, we use the simple but common approach of checking the maximum relative
difference between values for states in successive iterations,
but more sophisticated approaches have been devised for TSGs~\cite{KKKW18}.

\iffalse
\begin{center}
$
\mathlarger{
x_s^k = 
\begin{cases}
r(s) + \max\limits_{a \in A(s)} \left( \sum_{s'\in S} \delta(s,a)(s') \cdot x_{s'}^{k-1} + r(s,a) \right) & \mbox{if } s \in S_{C}  \\ %\in S_{\Diamond} \\
r(s) + \min\limits_{a \in A(s)} \left( \sum_{s'\in S} \delta(s,a)(s') \cdot x_{s'}^{k-1} + r(s,a) \right) & \mbox{otherwise} %s \in S_{N \setminus C} %\in S_{\Box}
\end{cases}
}
$
\end{center}
\fi
% ----------------------------------------------------------------------------------------------------------------

\subsection{Binary Decision Diagrams}\label{sec:bg_bdds}

A \emph{binary decision diagram} (BDD)~\cite{Bry86} is a rooted, directed acyclic graph
used to provide a compact representation of a Boolean function over a particular set of Boolean variables.
A BDD $\dd{b}$ over $n$ Boolean variables $\ddvars{x}=(\dd{x}_1,\dots,\dd{x}_n)$
represents a function $\ddfunc{b}:\Bset^n\ra\Bset$.
BDDs have two types of nodes:
(i) \emph{non-terminal} nodes,
which are labelled with a variable $\dd{x}_i$,
and whose outgoing edges are labelled 1 (``then'') and 0 (``else'');
and (ii) \emph{terminal} (leaf) nodes, labelled with 0 or 1.
For a valuation $\valn{v}=(v_1,\dots,v_n)\in\Bset^n$ of $\ddvars{x}$, the value of $f_\dd{b}(\valn{v})$
can be found by traversing the BDD $\dd{b}$ from its root to a terminal node,
taking at each non-terminal node the edge matching the value $v_i$ for its variable $\dd{x}_i$.
The value of $f_\dd{b}(\valn{v})$ is taken as the value of the terminal node that is reached.

By requiring that variables are ordered, from the root node downwards,
and by storing the graph in reduced form (merging isomorphic subgraphs,
and removing redundant nodes), BDDs can represent structured Boolean functions
very compactly and can be manipulated efficiently,
i.e., with operations whose complexity is proportional to the number
of nodes in the graph rather than the size of the function.
This includes all standard Boolean operators, for example,
$\ddop{Or}(\dd{b}_1,\dd{b}_2)$, which returns the BDD representing the function $\ddfuncs{b}{1}\lor\ddfuncs{b}{2}$.
We also use $\ddop{And}(\dd{b}_1,\dd{b}_2)$ and $\ddop{Not}(\dd{b})$, defined analogously.

\emph{Multi-terminal BDDs} (MTBDDs)~\cite{CMZ+93}, which are also sometimes known
as algebraic decision diagrams (ADDs)~\cite{BFG+97}, generalise BDDs
by allowing terminal nodes to be labelled with values from an arbitrary set $D$. 
Hence, they represent functions of the form $f:\Bset^n\ra D$.
% from a set of Boolean variables to an arbitrary set $D$.
Typically, we are interested in real-valued functions
and so an MTBDD $\dd{m}$ over $n$ Boolean variables $\ddvars{x}=(\dd{x}_1,\dots,\dd{x}_n)$
represents a function $\ddfunc{m}:\Bset^n\ra\Rset$.
Like for BDDs, a variety of useful operators for MTBDDs can be implemented.
In particular, we use:

\begin{itemize}
\item
$\ddop{Apply}(op,\dd{m}_1,\dd{m}_2)$, where $op$ is a binary operation over the reals:
returns the MTBDD representing the function $\ddfuncs{m}{1}\,op\,\ddfuncs{m}{2}$.
\item
$\ddop{IfThenElse}(\dd{b},\dd{m}_1,\dd{m}_2)$, where $\dd{b}$ is a BDD and $\dd{m}_1,\dd{m}_2$ are MTBDDs:
returns the MTBDD for the function with value $\ddfuncs{m}{1}$ if $\ddfunc{b}$ is $\true$ and $\ddfuncs{m}{2}$ otherwise.
\item
$\ddop{Const}(c)$, where $c\in\Rset$:
returns the MTBDD representing the constant function with value $c$.
\item
$\ddop{Abstract}(op,\ddvars{y},\dd{m})$, where $op$ is a commutative and associative binary operation over the reals
(here, we often use $\min$ or $\max$)
and $\ddvars{y}\subset\ddvars{x}$ is a subset of the variables of $\dd{m}$:
returns an MTBDD over variables $\ddvars{x}\backslash\ddvars{y}$ representing
the result of abstracting all the variables in $\ddvars{y}$ from $\dd{m}$
by applying $op$ over all possible values taken by the variables in $\ddvars{y}$.
% For example, Abstract(+, (x1 ), M) would give the MTBDD representing the function fM|x1 =0 + fM|x1=1 and Abstract(×, (x1, x2), M) would give the MTBDD representing the func- tion fM|x1 =0,x2 =0 × fM|x1 =0,x2 =1 × fM|x1 =1,x2 =0 × fM|x1 =1,x2 =1 . In the latter, M|x1 =b1 ,x2 =b2 is equivalent to (M|x1=b1)|x2=b2.
% \item
% $\ddop{Restrict}(\dd{b},\dd{m})$, where $\dd{m}$ is an MTBDD over variables and $\ddvars{x}$
% and $\dd{b}$ is a BDD over variables $\ddvars{y}\subset\ddvars{x}$ representing a \emph{cube},
% i.e., such that $\ddfunc{b}=1$ for precisely one valuation $\valn{v}$ of $\ddvars{y}$:
% returns the MTBDD over variables $\ddvars{x}\backslash\ddvars{y}$
% representing the function $\ddfunc{m}$ restrict to valuation $\valn{v}$ for $\ddvars{y}$.
\end{itemize}
BDDs were popularised thanks to the success of \emph{symbolic model checking}~\cite{BCM+90,McM93},
which uses them to provide an efficient and scalable implementation of
model checking, for example of the temporal logic CTL on labelled transition systems.
Assume that we have an encoding $\encfunc{S}:S\ra\Bset^k$
of the state space $S$ of a transition system into $k$ Boolean variables.
We can represent a subset $S'\subseteq S$ as a BDD,
by using it to encode the characteristic function $\chi_{S'}:S\ra\Bset$.
A transition relation $\ra\,\subseteq S\times S$
can be represented similarly as a BDD over 2 sets of $k$ Boolean variables,
i.e., by a BDD $\dd{b}$ where $\ddfunc{b}(\encfunc{S}(s),\encfunc{S}(s'))=1$
if and only if $(s,s')\in\,\ra$.
The key operations for model checking such as (pre or post) image computation
can be performed efficiently on these BDD representations.

Symbolic implementations of probabilistic model checking~\cite{BCHG+97,Par02}
build on the fact that real-value vectors and matrices can be
represented as MTBDDs in similar fashion.
A key operation used in the numerical computation required for probabilistic model checking
(i.e., for value iteration)
is matrix-vector multiplication, which can be performed symbolically~\cite{CMZ+93,CFM+93}:

\begin{itemize}
\item
$\ddop{MVMult}(\dd{m},\dd{v})$, where $\dd{m}$ is an MTBDD over variables $\ddvars{x},\ddvars{y}$
representing a matrix $\matr{M}$ and $\dd{v}$ is an MTBDD over variables $\ddvars{x}$ representing a vector $\matr{v}$:
returns the MTBDD over variables $\ddvars{x}$ representing the vector $\matr{M}\matr{v}$.
\end{itemize}

\section{Symbolic Model Checking for Stochastic Games}\label{sec:smc}

We now describe a symbolic implementation for the
representation, construction and verification of TSGs.

\subsection{Symbolic Representation and Construction of TSGs}

We begin by discussing how to represent TSGs symbolically, as MTBDDs.
The key components of a TSG, as required to perform model checking,
are the transition probability function $\delta:S\times\Act\ra\Dist(S)$
and the partition $(S_i)_{i \in N}$ of the state space amongst players.
We consider two different symbolic encodings,
one which represents $\delta$ and $(S_i)_{i}$ separately,
and one which uses a single MTBDD.

For the first, we can use the standard approach for MDPs~\cite{Bai98,dAKN+00}, 
which considers $\delta:S\times\Act\ra\Dist(S)$ as a function $\delta':S\times\Act\times S\ra[0,1]$
in the obvious way, i.e., for states $s,s'$ and action $a$,
we have $\delta'(s,a,s')=\delta(s,a)(s')$.
Then, given an encoding $\encfunc{S}:S\ra\Bset^k$ of the state space into $k$ Boolean variables,
and an encoding $\encfunc{\Act}:\Act\ra\Bset^l$ of the action set into $l$ Boolean variables,
$\delta$ can be represented by an MTBDD over $2k+l$ variables.
Reusing the same encoding $\encfunc{S}$,
each set $S_i$ is represented by a BDD over $k$ variables.

For the second encoding,
we assume that the TSG is represented by a single function
$\delta'':N\times S\times\Act\times S\ra[0,1]$
such that, for player $i$, states $s,s'$ and action $a$,
$\delta''(i,s,a,s')$ equals $\delta(s,a)(s')$ if $s\in S_i$ and 0 otherwise.
Given encodings $\encfunc{S}:S\ra\Bset^k$ and $\encfunc{\Act}:\Act\ra\Bset^l$ as above,
plus an encoding $\encfunc{N}:N\ra\Bset^m$ of the player set,
we can represent the TSG as an MTBDD over $2k+l+m$ Boolean variables.
In our experiments, we found minimal difference between the two encodings,
in terms of the size of storage for $\delta$, but the first option
incurs some additional overhead relating to the representation of the sets~$S_i$.
Hence, in this paper, we focus on the second, single-MTBDD encoding.

From now on, we will assume the use of variables
$\ddvars{x}=(\dd{x}_1,\dots,\dd{x}_k)$ and $\ddvars{y}=(\dd{y}_1,\dots,\dd{y}_k)$
to encode the state space $S$ (both $\dd{x}$ and $\dd{y}$ are used
when representing the transition function;
only one, usually $\ddvars{x}$, is needed when representing a subset of $S$
or a real-valued vector indexed over $S$).
We will use variables $\ddvars{z}=(\dd{z}_1,\dots,\dd{z}_l)$ to encode actions
and variables $\ddvars{w}=(\dd{w}_1,\dots,\dd{w}_m)$ to encode players.

\begin{figure}[t]
\begin{center}

$
\begin{array}{ccc}

\raisebox{.2\height}{
\hspace{0.4cm}
\begin{tikzpicture}[->,>=stealth',shorten >=1pt,auto,node distance=2.8cm, semithick, scale=.45]
  \tikzstyle{every state}=[draw=black,text=black, initial text=]

\small
   
\node[state,diamond,minimum width=0.75cm,minimum height=0.75cm] (S)at(0,-1.5) (s0) {$s_0$}; 
\node[state,rectangle,minimum width=0.75cm,minimum height=0.75cm] (S)at(5,0) (s1) {$s_1$};
\node[state,diamond,minimum width=0.75cm,minimum height=0.75cm] (S)at(5,-3) (s2) {$s_2$}; 
\node[circle,scale= 0.3,fill=black] (S)at(3,-1.5) (a) {}; 
\node[circle,scale= 0.3,fill=black,label={right:{$a$}}] (S)at(7.5,-1) (b) {}; 

\path [-] (s0.east) edge node [pos = 0.5] {$b$} (a.center);
\path [-] (s1.east) edge node [pos = 0.7] {} (b.center);
\path [->] (a) edge node [pos = 0.5] {$0.9$} (s1.west);
\path [->] (a) edge node [pos = 0.5, swap] {$0.1$} (s2.west);
\path [->] (b.center) edge node [pos = 0.5] {$0.9$} (s2.east);

\path (b.center) edge[pos = 0.35, out=45,in=15, looseness=1.5, swap] node {0.1} (s1.east);
\path (s0) edge[out=60, in=120, looseness=5] node [above] {$1\ \ a$} (s0);
\path (s2) edge[out=300, in=240, looseness=5] node {$1\ \ a$} (s2);

\end{tikzpicture} 
} 

\hspace{-5.5cm}

\raisebox{-2.0cm}{
\begin{tabular}{|c|c||c|c||c|c||c|}
\hline
Play. & $w_1,w_2$ & Act. & $z_1$ & $s\ra s'$ & $x_1,x_2\,\,y_1,y_2$ & Prob. \\
\hline
1 & 1,0 &$a$ & 0 & $s_0\ra s_0$ & 0,0 0,0 & $1$ \\
1 & 1,0 &$b$ & 1 & $s_0\ra s_1$ & 0,0 0,1 & $0.9$ \\
1 & 1,0 &$b$ & 1 & $s_0\ra s_2$ & 0,0 1,0 & $0.1$ \\
2 & 0,1 &$a$ & 0 & $s_1\ra s_1$ & 0,1 0,1 & $0.1$ \\
2 & 0,1 &$a$ & 0 & $s_1\ra s_2$ & 0,1 1,0 & $0.9$ \\
1 & 1,0 &$a$ & 0 & $s_2\ra s_2$ & 1,0 1,0 & $1$ \\
\hline
\end{tabular}
}

% \bordermatrix{ 
% w ~ z ~ {x|y} & s_0 & s_1 & s_2 \cr
% \mbox{1 ~F}~ s_0  & 1 & 0 & 0 \cr \vspace{0.15cm} \cr
% \mbox{1 ~T}~ & 0 & 0.9 & 0.1 \cr \vspace{0.15cm} \cr	
% \mbox{2 ~F}~ s_1  & 0 & 0.1 & 0.9 \cr \vspace{0.15cm} \cr
% \mbox{1 ~F}~ s_2 & 0 & 0 & 1
% }

&
% \rightarrow
&

\hspace{-0.2cm}

\raisebox{-.5\height}{
\begin{forest}
  BDT
  [$w_1$, name=gw11 
        [$w_2$, name=gw21,
            [$z_1$, edge=solid,
                [$x_1$, 
                    [$y_1$, 
                        [$x_2$, 
                            [$y_2$, edge=solid,
                                [0.1, name=gv0p1, edge=solid, node options={right of=gw21}]
                            ]
                        ]
                        [$x_2$,
                            [$y_2$, edge=solid,
                                [0.9, name=gv0p9, node options={right of=gw21}]
                            ]
                        ]
                    ]
                ]
            ]
        ]
        [$w_2$, name=gw22,
            [$z_1$, 
                [$x_1$, edge=solid,
                    [$y_1$
                        [$x_2$, edge=solid,
                            [$y_2$, node options={draw,circle}, tikz={\draw [dashed] () [] to (gv0p1.45);}
                            ]
                        ]
                        [$x_2$, edge=dashed,
                            [$y_2$, node options={draw,circle}, tikz={\draw [solid] () [] to (gv0p9.45);}
                            ]
                        ]
                    ]
                ]
                [$x_1$, edge=dashed,
                    [$y_1$
                        [$x_2$, name=gx25,
                            [$y_2$
                                [1]
                            ]
                        ]
                    ]
                    [$y_1$, node options={draw,circle}, tikz={\draw [solid] () [] to (gx25.north east);}
                    ]
                ]
            ]
        ]
  ]
\end{forest}
}

\end{array}
$
\end{center}
\vspace{-0.2cm}
\caption{A TSG with its MTBDD representation and an explanation of the encoding.}
\label{fig:tsg}
\end{figure}
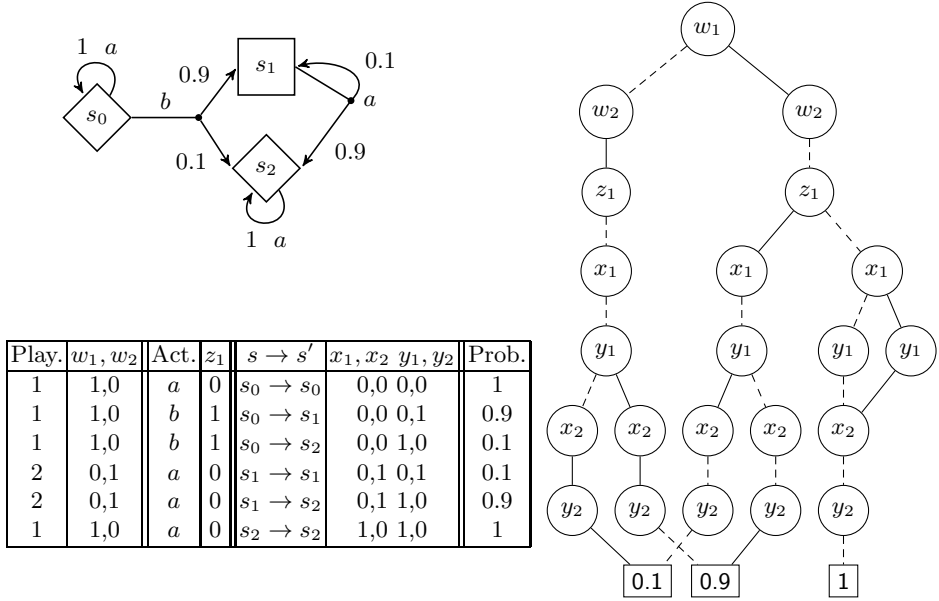

\begin{examp}
\figref{fig:tsg} shows a simple TSG with 2 players and its symbolic representation,
using the second (single MTBDD) encoding described above.
Top left is the TSG, in which player 1 states are drawn as diamonds
and player 2 states as squares.
Below that is a table explaining the representation:
the details of each transition in the TSG
and how it is encoded into Boolean variables.

For players, we use a one-hot encoding to two Boolean variables $(\dd{w}_1,\dd{w}_2)$,
i.e., $\encfunc{N}(1)=(1,0)$ and $\encfunc{N}(2)=(0,1)$.
For the (two) actions, we use just a single variable $\dd{z}_1$,
where $\encfunc{\Act}(a)=(0)$ and $\encfunc{\Act}(b)=(1)$.
The state space $S$ is encoded with 2 variables
using the usual binary encoding of the integer index $i$ of each state $s_i$.
In a transition, variables $(\dd{x}_1,\dd{x}_2)$ and $(\dd{y}_1,\dd{y}_2)$
represent the source and destination states, respectively.

To the right of the figure is the MTBDD representation.
The 1 (``then'') and 0 (``else'') edges
from each non-terminal node are drawn as solid and dashed lines, respectively.
The zero terminal and edges to it are omitted for clarity.
Each row of the table corresponds to a unique path through the MTBDD.
The variable order used places $\ddvars{w}$ and $\ddvars{z}$ first,
followed by $\ddvars{x}$ and $\ddvars{y}$, where, as usual in symbolic model checking,
the variables in the latter two are interleaved.
\end{examp}

In order to be effective in practice, symbolic representations of TSGs need to
be \emph{constructed} in an efficient manner.
In the context of this paper, we work with games that are described
in the modelling language of the PRISM-games tool~\cite{KNPS20},
which is inspired by the Reactive Modules formalism~\cite{AH99},
proposed for specifying concurrent, multi-component systems.

We omit full details here, but note that this can be done by
extending the existing approach used for the symbolic implementation
of model checking for simpler probabilistic models in PRISM~\cite{KNP11}.
The basics for model construction from the PRISM modelling language can be
found in \cite{Par02}. The key idea is to construct the MTBDD in a compositional
fashion, based on the structure of the model description.
We also note that the second MTBDD encoding, building a single MTBDD,
is better suited for this task, since it facilitates the detection of modelling
errors (such as multiple players controlling actions in the same state).

%------------------------------------------------------------------------------------------

\subsection{Symbolic Model Checking of TSGs}

Next, we describe a symbolic approach to performing probabilistic model checking of TSGs.
We focus here on the PRISM-games logic described in \sectref{sec:bg_mc}.
Essentially, since this is a branching-time logic,
the model checking problem for a TSG $\game$ and a formula $\phi$
amounts to determining the set $\Sat(\phi)=\{s\in S\,|\,s\models\phi\}$.
Furthermore, this set is computed in a recursive fashion,
following the structure of the parse tree of the formula $\phi$.

In a symbolic setting, the set $\Sat(\phi)$ will be represented as a BDD.
The propositional fragment of the logic is treated in the usual
way for symbolic model checking~\cite{BCM+90,McM93}, using standard BDD implementations of Boolean operators.
The key parts of the model checking algorithm are those for the $\probopP$ and $\rewopR$ operators.
In particular, we need to compute an MTBDD representing the real-valued vector
of probability or expected reward values for each state $s$.

Computing these values can be done in a variety of ways.
Here, we use \emph{value iteration},
since iterative methods are known to be typically better suited to symbolic implementation%
%for simpler probabilistic models such as Markov chains and Markov decision processes
~\cite{BFG+97,BCHG+97,Par02}.
This is because it requires minimal changes to be made to the representation
of the model during solution, which could cause a blow-up in storage size
due to the introduction of irregularities.
This means that some alternative methods for solving stochastic games such as quadratic programming
are unlikely to be well suited to a symbolic implementation.

To simplify presentation, we restrict our attention to computing reachability probabilities,
assuming that they are maximised by a coalition of players $C$ (and minimised by $N\backslash C$).
The process for other computations, such as expected reward values, is similar.
In other words, we consider the construction of an MTBDD $\dd{sol}$ representing a vector $\vect{sol}$
indexed over $S$ with $sol(s) = \val(s,\coalition{C} \probop{\max=?}{\future\mathit{target}})$
for some atomic proposition $\mathit{target}$ labelling the states to be reached
(see \sectref{sec:bg_mc}).

\begin{algorithm}[!t]
\caption{Value iteration and strategy synthesis for reachability probabilities}
%\hspace*{\algorithmicindent} 
\textbf{Input:} BDD $\dd{target}$ (over variables $\ddvars{x}$) for set of target states \\
%\hspace*{\algorithmicindent} 
\textbf{Output:} MTBDD $\dd{sol}$ (over variables $\ddvars{x}$) giving the probability from each state,
and a BDD $\dd{strat}$ (over variables $\ddvars{x},\ddvars{z}'$) representing an optimal strategy
\begin{algorithmic}[1]
\PROCEDURE{\alg{ProbReach}($\dd{target}$)}
\STATE $\dd{S_0}\leftarrow \alg{Prob0}(\dd{target})$
\STATE $\dd{S_1}\leftarrow \alg{Prob1}(\dd{target})$
\STATE $\dd{S_?}\leftarrow \alg{Not}(\alg{Apply}(\lor, \dd{S_1}, \dd{S_0}))$
\STATE $\dd{trans_?}\leftarrow \ddop{Apply}(\times,\dd{trans},\dd{S_?})$
\STATE $\dd{sol}\leftarrow\dd{S_1}$ ; $done \leftarrow$ \FALSE
\WHILE{$\neg done$} 
    % \STATE $\ddop{ReplaceVars}$
    \STATE $\dd{tmp}\leftarrow\dd{sol}$
    \STATE $\dd{sol}\leftarrow\alg{MVMult}(\dd{trans_?}, \dd{sol})$
    \STATE $\dd{p}_1\leftarrow \alg{Or}(\{\ddop{Cube}(\enc{N}{i},\ddvars{w})\,:\,i\in N\})$
    \STATE $\dd{p}_2\leftarrow \alg{Or}(\{\ddop{Cube}(\enc{N}{i},\ddvars{w})\,:\,i\in N\backslash C\})$
    \STATE $\dd{sol_1}\leftarrow\alg{Abstract}(\max, \ddop{Abstract}(+, \ddop{apply}(\times, \dd{sol}, \dd{p_1}), \ddvars{w}), \ddvars{z})$
    \STATE $\dd{sol_2}\leftarrow\alg{Abstract}(\min, \ddop{Abstract}(+, \ddop{apply}(\times, \dd{sol}, \dd{p_2}), \ddvars{w}), \ddvars{z})$
    % \STATE $\dd{sol_1}\leftarrow\alg{Abstract}(\max, \ddop{Restrict}(\dd{sol}, \dd{p_1}), \ddvars{z})$
    % \STATE $\dd{sol_2}\leftarrow\alg{Abstract}(\min, \ddop{Restrict}(\dd{sol}, \dd{p_2}), \ddvars{z})$
    \STATE $\dd{sol}\leftarrow\alg{Apply}(+, \dd{sol}_1, \dd{sol}_2)$
    \STATE $\dd{sol}\leftarrow\alg{IfThenElse}(\dd{S_1}, \ddop{Const}(1), \dd{sol})$
    \STATE $\mathit{done}\leftarrow\alg{SupNorm}(\dd{sol}, \dd{tmp}) < \varepsilon$
\ENDWHILE
\STATE $\dd{strat}\leftarrow \ddop{Apply}(\approx,\alg{MVMult}(\dd{trans_?}, \dd{sol}),\dd{sol})$
\STATE $\dd{strat}\leftarrow\alg{Abstract}(\lor, \dd{strat}, \ddvars{w})$
\STATE $\dd{strat}\leftarrow \ddop{ReplaceVars}(\dd{strat},\ddvars{z},\ddvars{z}')$
\RETURN $\dd{sol},\dd{strat}$
\ENDPROCEDURE
\end{algorithmic}
\label{algo:valiter}
\end{algorithm}

\algoref{algo:valiter} shows an MTBDD implementation
that performs both the numerical solution, using value iteration,
and synthesis of an optimal strategy.
The input is a BDD $\dd{target}$ representing the target set $\Sat(\mathit{target})$,
and we assume that MTBDD $\dd{trans}$ encodes the TSG.
$\ddop{Prob0}$ and $\ddop{Prob1}$ are BDD-based implementations of
the precomputation algorithms~\cite{AHK02} for finding states with probability 0 and 1;
we omit the details and focus on the numerical part.

The key part of value iteration can be done using matrix multiplication,
treating the TSG as a non-square matrix with rows over $N\times\Act\times S$ and columns over $S$,
followed by maximising and minimising over action choices for players in $C$ and $N\backslash C$, respectively.
Function $\ddop{Cube}(\valn{v},\ddvars{w})$ builds a \emph{cube},
i.e., a BDD $\dd{b}$ over variables $\ddvars{w}$
such that $\ddfunc{b}=1$ for precisely one valuation $\valn{v}$ of $\ddvars{w}$.
We check termination of value iteration using a function $\ddop{SupNorm}$
which performs a pointwise calculation of the relative difference for pairs of elements in two vectors
represented as MTBDDs and returns the maximum difference.
This is compared against a pre-specified convergence criterion threshold $\varepsilon\in\Rset_{>0}$.

\startpara{Strategy synthesis}
Lines 18-20 compute an optimal strategy,
where $\approx$ represents an approximate equality check to the same level of accuracy
as the convergence check (i.e., relative difference less than $\varepsilon$),
and $\ddvars{z}'$ is a fresh copy of the variables $\ddvars{z}$ that encode actions,
but appearing after $\ddvars{x}$ and $\ddvars{y}$ in the variable ordering.
The result is a BDD $\dd{strat}$ over variables $\ddvars{x}$ and $\ddvars{z}$,
representing an optimal strategy: for any state $s$, we traverse the top part of the BDD
by following valuation $\enc{S}{s}$. Any path (we allow multiple) from that node to the
1 terminal represents an optimal action $a$ in that state (read from its encoding $\enc{\Act}{a}$).
We leave as future work the possibility of selecting single optimal actions  for states
in a way that further reduces the size of the strategy representation.

\startpara{Nash equilibria}
Lastly, we briefly sketch how our symbolic model checking implementation also extends to nonzero-sum formulas,
i.e., the synthesis of (social welfare) Nash equilibria.
The process is again based on value iteration but, as mentioned in \sectref{sec:bg_mc},
this is simpler for TSGs than the CSG-based algorithm of~\cite{KNPS21}.
Essentially we adapt \algoref{algo:valiter},
first maximising for individual coalitions, as in the existing value iteration loop,
then selecting all actions that are optimal, as in the strategy synthesis part,
and then further maximising those choices over the sum of values for all players.
The latter part means that we maintain a solution vector for each player as MTBDDs during the process.
For the 2-coalition case, part of the computation reduces to symbolic model
checking for MDPs, where we can reuse existing implementations.

\section{Case Studies and Experimental Results}\label{sec:exp}

We have developed a symbolic implementation of model checking for TSGs within PRISM-games~\cite{KNPS20},
leveraging parts of PRISM's existing symbolic engines for other models (Markov chains and Markov decision processes).
This builds upon the CUDD decision diagram library by Fabio Somenzi, which supports both BDDs and MTBDDs,
and a Java wrapper contained within PRISM which extends this library.
Our experiments were carried out using a 2.10GHz Intel Xeon Gold
with 16 GB maximum heap space for Java.

\subsection{TSG Benchmarks}

In order to evaluate the approach, we first present a set of benchmark TSG models.
We have collated these and added them to the PRISM Benchmark Suite~\cite{KNP12b},
which provides a selection of probabilistic models and associated properties
for performing model checking.
To facilitate benchmarking, most models and properties are parameterised,
allowing a wide range of model checking instances to be considered.
Python scripts are also included to automate the process of selecting
and executing instances, and for extracting information from tool logs.

The benchmarks are listed below: % and more details are available at~\cite{www}.

\begin{itemize}
\item \emph{avoid}: a TSG example from~\cite{CKWW20}
modelling a game between an intruder and an observer in a grid-world (also used in \cite{KRSW20});
\item \emph{dice}: a simple 2-player dice game TSG distributed with PRISM-games;
\item \emph{hallway{\uu}human}:
a TSG variant (from \cite{CKWW20}) of a standard benchmark from the AI literature~\cite{LCK95}
modelling a robot moving through a hallway environment which is both probabilistic and adversarial (also used in \cite{KRSW20});
\item \emph{investors}: the futures market investor TSG  example from \cite{MM07},
adapted and extended to more investors;
% $\coalition{investor_1}\probop{\max=?}{\futureop \, \mathsf{done} \land v>v_{max}/2}$
%\item \emph{repudation}:
%a TPTG (turn-based probabilistic timed game) model of Markowitch \& Roggeman’s non-repudiation protocol for information transfer~\cite{KNP19};
\item \emph{safe{\uu}nav}: a TSG modelling safe navigation in a human-robot system, from~\cite{JJK+18}.
\item \emph{task{\uu}graph}: an extended version of the task-graph scheduling problem with faulty processors from \cite{KNP19},
converted from a (turn-based) probabilistic timed game to a TSG using the digital clocks translation of~\cite{KNP19}.
\end{itemize}

\begin{table}[t!]
\centering
{\small
\begin{tabular}{|c|c||c|r|r|r|r|r|} \hline
\multicolumn{1}{|c|}{Case study} & 
\multicolumn{1}{c||}{Param.} & 
\multicolumn{1}{c|}{Players} & \multicolumn{1}{c|}{States} &  %\multicolumn{1}{c|}{Transitions} & 
\multicolumn{1}{c|}{MTBDD} &
\multicolumn{2}{c|}{Constr. time(s)} \\ \cline{6-7} \multicolumn{1}{|c|}{[parameters]} &
\multicolumn{1}{c||}{values} & & & \multicolumn{1}{c|}{nodes} & \multicolumn{1}{c|}{Symbolic} & \multicolumn{1}{c|}{Explicit}
 \\ \hline \hline	
\multirow{3}{*}{\shortstack[c]{\emph{avoid} \\ $[\mathtt{X{\uu}MAX},\mathtt{Y{\uu}MAX}]$}}
& 10,10 & \multirow{3}{*}{2} &   106,524 &%   297,790 
19,298 & \best{0.2} &  1.6 \\
& 15,15 &                    &   480,464 &% 1,380,660 
36,178 & \best{0.4} &  6.4 \\
& 20,20 &                    & 1,436,404 &% 4,182,830 
69,407 & \best{1.0} & 18.8 \\
\hline\hline
\multirow{3}{*}{\shortstack[c]{\emph{dice} \\ $[\mathtt{N}]$}}
& 10 & \multirow{3}{*}{2} &  5,755 &%    7464 
1,717 & \best{0.02} & 0.2 \\
& 25 &                    & 34,645 &%  25,764 
4,046 & \best{0.04} & 0.5 \\
& 50 &                    & 136,795 &% 145,464 
7,958 & \best{0.09} & 1.5 \\
\hline\hline
\multirow{3}{*}{\shortstack[c]{\ \emph{hallway{\uu}human\ } \\ $[\mathtt{X{\uu}MAX},\mathtt{Y{\uu}MAX}]$}}
&   5,5 & \multirow{3}{*}{2} &  25,000 &%   108,200 
1,334 & \best{0.03} & 0.6 \\
&   8,8 &                    & 163,840 &%   727,040 
1,234 & \best{0.04} & 2.8 \\
& 10,10 &                    & 400,000 &% 1,788,800 
1,752 & \best{0.07} & 6.7 \\
\hline\hline
\multirow{6}{*}{\shortstack[c]{\emph{investors} \\ $[\mathtt{N},\mathtt{vmax}]$}}
& 2,10 & \multirow{3}{*}{3} &   172,240 &%   373,669 
5,846 & \best{0.04} &  2.0 \\
& 2,20 &                    &   568,790 &% 1,247,069 
11,325 & \best{0.06} &  6.5 \\
& 2,40 &                    & 2,041,690 &% 4,503,469 
22,191 &  \best{0.1} & 23.4 \\
\cline{2-7}
& 3,10 & \multirow{3}{*}{4} &  1,229,001 &%  2,622,435 
7,434 & \best{0.06} &  13.7 \\
& 3,20 &                    &  4,058,751 &%  8,712,860 
12,913 &  \best{0.1} &  48.5 \\
& 3,40 &                    & 14,569,251 &% 31,383,810 
23,779 &  \best{0.2} & memout \\
\hline\hline
\multirow{4}{*}{\shortstack[c]{\emph{safe{\uu}nav} \\ $[\mathtt{N},\mathtt{feat}]$}}
& 8,D & \multirow{4}{*}{2} &     2,592,845 &%  6,698,509 
28,008 & \best{1.0} &  1,602 \\
& 8,C &                    &     5,078,029 &% 13,269,872 
44,973 & \best{1.7} &  4,588 \\
& 8,B &                    &     8,732,493 &% 22,957,256 
67,735 & \best{2.7} &  10,010 \\
& 8,A &                    &    17,052,941 &% 45,581,400 
118,262 & \best{4.8} & memout \\
\hline\hline
\multirow{6}{*}{\shortstack[c]{\emph{task{\uu}graph} \\ $[\mathtt{N},\mathtt{k1},\mathtt{k2}]$}} 
& \ \ 6,10,10\ \ & \multirow{3}{*}{2} & 467,638 &%  8,121,271 
19,881 & \best{0.6} &   6.7 \\
& 6,15,15 &                    & 1,010,318 &% 17,690,891 
22,350 & \best{1.0} &  13.8 \\
& 6,20,20 &                    & 1,759,348 &% 30,937,011 
22,350 & \best{1.8} &  25.1 \\
\cline{2-7}
& 9,10,10 & \multirow{3}{*}{2} & 2,567,638 &%  8,121,271 
36,014 & \best{1.4} &  46.5 \\
& 9,15,15 &                    & 5,533,288 &% 17,690,891 
36,745 & \best{2.8} & 100.0 \\
& 9,20,20 &                    & 9,6231,38 &% 30,937,011 
39,349 & \best{4.6} & 169.2 \\
\hline
\end{tabular}
}
\vspace{0.2cm}
\caption{Model building statistics for the TSG case studies.}
\label{tab:model_build}
\end{table}

% \gabriel{for some, the memory limit for the explicit need to be higher.}
\begin{table}[t!]
\centering
{\small
\begin{tabular}{|c|c||r|r|r|r||r|r|r|r|} \hline
\multicolumn{1}{|c|}{Case study} & 
\multicolumn{1}{c||}{Param.} & 
\multicolumn{8}{c|}{Verification time and strategy memory} \\ \cline{3-10} 
\multicolumn{1}{|c|}{{[parameters]}} & 
\multicolumn{1}{c||}{values} & \multicolumn{4}{c||}{Symbolic} & \multicolumn{4}{c|}{Explicit} \\ \cline{3-10}
Property (type) & & Qual. & Quant. & \multicolumn{1}{c|}{Total} & Strat. & Qual.& Quant. & \multicolumn{1}{c|}{Total} & Strat. \\
 & & \multicolumn{1}{c|}{(s)} & \multicolumn{1}{c|}{(s)} & \multicolumn{1}{c|}{(s)} & \multicolumn{1}{c||}{(MB)} & \multicolumn{1}{c|}{(s)} & \multicolumn{1}{c|}{(s)} & \multicolumn{1}{c|}{(s)} & \multicolumn{1}{c|}{(MB)} \\ \hline \hline	
\multirow{3}{*}{\shortstack[c]{\emph{avoid} \\ $[\mathtt{X{\uu}MAX},\mathtt{Y{\uu}MAX}]$ \\
\emph{exit} ($\probop{}{\futureop}$)
}}
& 10,10 &  \best{4.4} &  4.0 &   \best{8.4} & \best{0.1} &  33.2 &  \best{0.9} &  34.2 & 0.4 \\
& 15,15 & \best{20.4} & 23.0 &  \best{43.6} & \best{0.2} & 407.2 &  \best{9.1} & 416.3 & 1.8 \\
& 20,20 & \best{77.7} & 82.8 & \best{161.2} & \best{0.4} & 1,544 & \best{14.5} & 1,558 & 5.5 \\
\hline\hline
\multirow{3}{*}{\shortstack[c]{\emph{avoid} \\ $[\mathtt{X{\uu}MAX},\mathtt{Y{\uu}MAX}]$ \\
\emph{find} ($\probop{}{\futureop}$)}}
& 10,10 &   \best{8.3} &  4.0 &  \best{12.3} & \best{0.2} &  17.4 & \best{0.5} &  18.0 & 0.4 \\
& 15,15 &  \best{37.8} & 21.9 &  \best{60.0} & \best{0.4} & 224.9 & \best{3.7} & 228.7 & 1.8 \\
& 20,20 & \best{152.7} & 66.7 & \best{220.0} & \best{0.7} & 1,145 & \best{9.2} & 1,155 & 5.5 \\
\hline\hline
\multirow{3}{*}{\shortstack[c]{\emph{dice} \\ $[\mathtt{N}]$ \\
\emph{p1wins} ($\probop{}{\futureop}$)
}}
& 10 & \best{0.02} & \best{0.02} & \best{0.02} & \best{0.02} & 0.07 & 0.04 & 0.1 & 0.02 \\
& 25 &  \best{0.2} &  \best{0.2} &  \best{0.4} & \best{0.05} &  0.5 &   0.2 &  0.8 & 0.1 \\
& 50 &  \best{0.6} &  \best{0.5} &  \best{1.1} & \best{0.1} &  5.3 &   2.6 &  7.9 & 0.5 \\
\hline\hline
\multirow{3}{*}{\shortstack[c]{\ \emph{hallway{\uu}human\ } \\ $[\mathtt{X{\uu}MAX},\mathtt{Y{\uu}MAX}]$ \\
\emph{save} ($\probop{}{\futureop}$)}}
& 5,5   & \best{0.06} & - & \best{0.06} & - & 0.2 & - & 0.2 & - \\
& 8,8   &  \best{0.2} & - &  \best{0.2} & - & 2.0 & - & 2.0 & - \\
& 10,10 &  \best{0.6} & - &  \best{0.6} & - & 6.9 & - & 7.0 & - \\
\hline\hline
\multirow{6}{*}{\shortstack[c]{\emph{investors} \\ $[\mathtt{N},\mathtt{vmax}]$ \\
\emph{greater} ($\probop{}{\futureop}$)}}
& 2,10 & \best{0.04} &  \best{1.0} &  \best{1.1} & \best{0.06} &   3.1 &     3.1 &     6.4 &  0.7 \\
& 2,20 & \best{0.04} &  \best{5.4} &  \best{5.6} &  \best{0.1} &  19.7 &    22.3 &    42.2 &  2.2 \\
& 2,40 & \best{0.05} & \best{22.6} & \best{22.8} &  \best{0.2} &  27.8 &    97.1 &   125.3 &  7.8 \\
\cline{2-10}
& 3,10 &  \best{0.1} &  \best{3.4} &  \best{3.6} &  \best{0.2} &  27.6 &    30.9 &    58.9 &  4.7 \\
& 3,20 &  \best{0.1} & \best{16.0} & \best{16.2} &  \best{0.3} &  83.8 &   169.8 &   255.5 & 15.5 \\
& 3,40 &  \best{0.2} & \best{62.5} & \best{62.9} &  \best{0.4} &     - &       - &  memout &    - \\
\hline\hline
\multirow{4}{*}{\shortstack[c]{\emph{safe{\uu}nav} \\ $[\mathtt{N},\mathtt{feat}]$ \\
\emph{reach} ($\probop{}{\futureop}$)}}
& 8,D &  \best{12.7} &         4.1 &  \best{17.1} & \best{1.8} & 134.9 & \best{2.5} & 138.0  & 9.9 \\
& 8,C &  \best{26.1} &         7.6 &  \best{34.4} & \best{2.9} & 145.8 & \best{3.6} & 150.4  & 19.4 \\
& 8,B &  \best{48.8} &        12.0 &  \best{62.1} & \best{4.6} & 313.8 & \best{7.6} & 323.1  & 33.3 \\
& 8,A & \best{138.7} & \best{27.1} & \best{169.3} & \best{8.7} &     - &          - & memout & - \\
\hline\hline
\multirow{6}{*}{\shortstack[c]{\emph{task{\uu}graph} \\ $[\mathtt{N},\mathtt{k1},\mathtt{k2}]$ \\
\emph{time} ($\rewop{}{}{\futureop}$)}}
& \ \ 6,10,10\ \ & \best{0.8} & 116.8 &        117.7 & - &  14.3 &  \best{34.2} &  \best{48.9} & - \\
&        6,15,15 & \best{1.1} & 346.6 &        348.1 & - &  27.9 &  \best{63.4} &  \best{91.8} & - \\
&        6,20,20 & \best{1.4} & 826.5 &        828.6 & - &  52.9 & \best{116.8} & \best{170.8} & - \\
\cline{2-10}
&        9,10,10 & \best{4.1} & 1,117 &        1,122 & - &  90.9 & \best{179.3} & \best{271.5} & - \\
&        9,15,15 & \best{5.7} & 3,304 &        3,312 & - & 250.9 & \best{515.9} & \best{769.9} & - \\
&        9,20,20 & \best{7.8} & 6,624 &        6,636 & - & 660.2 & \best{1,268} & \best{1,934} & - \\
\hline
\end{tabular}
}
\vspace{0.2cm}
\caption{Statistics for TSG verification instances.}
\label{tab:model_check}
\end{table}

\subsection{Experimental Results}

\tabref{tab:model_build} shows statistics for a selection of TSG model instances that we use for our evaluation
(see~\cite{www,KNP12b} for more details).
We also give the time required to build a representation of the TSG, from its PRISM-games modelling language description,
either symbolically, as an MTBDD, or explicitly, as a sparse matrix, as done in the existing implementation of PRISM-games.
The faster time is in bold.
For the symbolic case, we also show the MTBDD size.
We see that the symbolic approach is considerably faster.
The explicit implementation of model construction (in Java) is not highly optimised
but the difference in performance is clear nonetheless.
For some instances, where the explicit engine took several hours, the symbolic one requires no more than a few seconds.

Secondly, \tabref{tab:model_check} shows the performance of model checking for the symbolic and explicit implementations
on a range of example properties for the benchmarks (again, see~\cite{www,KNP12b} for full details).
We break down the time required into qualitative analysis (graph-based precomputation)
and quantitative analysis (numerical solution with value iteration).
Again, the faster time is highlighted in bold.
We also show the total memory required to store the resulting optimal strategy in each case
(these are omitted for qualitative probabilistic reachability and expected reward,
since they are not yet included in the implementation).

Pre-computation has shown to be more efficient for all model/property combinations in the table,
and in some cases this is a decisive factor in terms of the overall model checking time. 
Results for value iteration generally favour the explicit engine,
although there are instances where the symbolic one performs better.    
In terms of representing optimal strategies, we see that the symbolic one
is more compact in all cases.

\section{Conclusions}\label{sec:conc}

We have presented a symbolic version of probabilistic model checking for
turn-based stochastic games, using BDDs and MTBDDs to implement model construction,
model checking (via value iteration) and optimal strategy synthesis.
There are some significant gains to be had, particularly in terms of model construction,
but also further improvements to be made.

Future work includes studying different encodings more thoroughly, as well as variable orderings.
There is also scope to investigate more efficient symbolic strategy representations.
Another possibly interesting extension is providing support for B\"uchi, co-B\"uchi and Rabin-chain objectives \cite{LM04}, where a symbolic implementation could also allow for better scalability.

\vskip0.5em

\startpara{Acknowledgements}
This project was funded by the ERC under the European Union’s Horizon 2020 research and innovation programme (\href{http://www.fun2model.org}{FUN2MODEL}, grant agreement No.~834115).

% \clearpage
\bibliographystyle{splncs04.bst}
\bibliography{bib}

% \clearpage
% \input{appendix}

\end{document}